# Structural Damage Identification Using Piezoelectric Impedance Measurement with Sparse Inverse Analysis

Pei Cao

Graduate Research Assistant and Ph.D. Candidate

Department of Mechanical Engineering

University of Connecticut

Storrs, CT 06269, USA

Qi Shuai

Assistant Professor

Department of Automotive Engineering

Chongqing University

Chongqing 400044, China

J. Tang[↑]

Professor

Department of Mechanical Engineering

University of Connecticut

Storrs, CT 06269, USA

Phone: (860) 486-5911, Email: jiong.tang@uconn.edu

[↑] Corresponding author

# Structural Damage Identification Using Piezoelectric Impedance Measurement with Sparse Inverse Analysis

Pei Cao[1], Shuai Qi[2], and J. Tang[1][↑]

1: Department of Mechanical Engineering, University of Connecticut

Storrs, CT USA 06269

2: Department of Automotive Engineering, Chongqing University

Chongqing 400044, PR China

ABSTRACT

The impedance/admittance measurements of a piezoelectric transducer bonded to or embedded in a host structure can be used as damage indicator. When a credible model of the healthy structure, such as the finite element model, is available, using the impedance/admittance change information as input, it is possible to identify both the location and severity of damage. The inverse analysis, however, may be under-determined as the number of unknowns in high-frequency analysis is usually large while available input information is limited. The fundamental challenge thus is how to find a small set of solutions that cover the true damage scenario. In this research we cast the damage identification problem into a multi-objective optimization framework to tackle this challenge. With damage locations and severities as unknown variables, one of the objective functions is the difference between impedance-based model prediction in the parametric space and the actual measurements. Considering that damage occurrence generally affects only a small number of elements, we choose the sparsity of the unknown variables as another objective function, deliberately, the $l_0$ norm. Subsequently, a multi-objective Dividing RECTangles (DIRECT) algorithm is developed to facilitate the inverse analysis where the sparsity is further emphasized by sigmoid transformation. As a deterministic technique, this approach yields results that are repeatable and conclusive. In addition, only one algorithmic parameter, the number of function evaluations, is needed. Numerical and experimental case studies demonstrate that the proposed framework is capable of obtaining high-quality damage identification solutions with limited measurement information.

---

[↑] Corresponding author

# 1. INTRODUCTION

Extensive research has been conducted on structural health monitoring (SHM) to protect structures from catastrophic failures. A structural health monitoring (SHM) system typically uses structural dynamic responses measured by sensors to elucidate the health condition. A traditional class of methods is vibration-based SHM, which analyzes and interprets measureable modal properties such as natural frequencies and mode shapes to determine damage locations and severities (Kim and Stubbs, 2003; Maity and Tripathy, 2005; Jiang et al, 2006; Jassim et al, 2013; Cao et al, 2017). However, in practical situations, normally only lower-order modes with large wavelengths can be realistically excited and measured; thus, these methods may not be sensitive to small-sized damage (Kim and Wang, 2014). Another well-known class of methods is based on wave propagation, which uses the change of transient wave upon its passage through damage site to infer damage occurrence (Michaels and Michaels, 2007; Harley and Moura, 2014). While these high-frequency methods may entertain high detection sensitivity, it is generally difficult to use the transient responses to identify damage accurately, especially to quantify the severity of damage (Cawley and Simonetii, 2005). Piezoelectric transducers are frequently used in wave propagation-based SHM. The two-way electro-mechanical coupling of piezoelectric transducers has also allowed them to be used in piezoelectric impedance- or admittance-based methods (Park et al, 2003; Min et al, 2012; Annamdas and Radhika, 2013; Shuai et al, 2017). In these methods, frequency-swept harmonic voltage excitations are applied, and stationary wave responses are induced and sensed. As such, the impedance of the structure is coupled with that of the piezoelectric transducer. The change of piezoelectric impedance signature with respect to that under the undamaged baseline can be used as damage indicator.

When a credible, first-principle-based baseline model, such as the finite element model, of the healthy structure is available, we may be able to identify both the fault location and severity by using vibration responses or impedance/admittance responses which are stationary (Xia and Hao, 2003; Jiang et al, 2006; Wang and Tang, 2008; Zhou and Zuo, 2012; Shuai et al, 2017). In such an inverse analysis, we typically divide the structure into a number of segments and assume that each segment in the model is susceptible of damage occurrence, i.e., certain property of each segment is an unknown. The inverse analysis uses the changes of stationary responses, such as natural frequencies, mode shapes, response amplitudes, impedances or admittances, as inputs. Indeed, a linearized sensitivity matrix can be derived that links the segment property change vector with respect to the response measurement change vector. While this appears to be an appealing formulation, there is a fundamental challenge, i.e., how to solve for the unknowns, i.e., damage locations and severities, from this seemingly straightforward relation. On one hand, the number of segments or the number of finite elements is generally very large, especially for SHM schemes utilizing high-frequency excitations such as those using piezoelectric impedance or admittance measurements where the mesh size must be smaller than the wavelength of the high-frequency responses. On the other hand, available measurement information is usually limited. For example, in impedance-based methods, only the impedance measurements around resonant peaks are truly sensitive to damage occurrence with high signal-to-noise ratio, and adding more measurement frequency points do not necessarily increase the row-rank of the sensitivity matrix. Therefore, the aforementioned inverse formulation may easily become under-determined (Kim and Wang 2014). Moreover, the inevitable measurement noise and modeling uncertainty further compound the difficulty (Shuai et al, 2017).

To avoid the direct inversion of the sensitivity matrix, alternatively, the problem of identifying damage location/severity in a finite element-based analysis can be cast into a global optimization formulation. Indeed, under the umbrella of optimization formulation, several global optimization techniques, such as particle swarm optimization (Begambre and Laier, 2009), differential evolution algorithm (Seyedpoor et al., 2015), genetic algorithm (Perera et al., 2010), and DIRECT algorithm (Cao et al, 2017), have been attempted, where possible property changes in all segments are treated as unknowns to be solved. One important feature of these optimization-based formulations is that usually only forward-analysis of the model is involved to facilitate comparison and minimization of the difference between model prediction in the parametric space and actual measurements. Nevertheless, given that the sensitivity matrix is under-determined, the fundamental mathematical challenge remains, i.e., there are in theory infinitely many solutions. Inevitably, many solutions that are different from the true damage scenario may be obtained. Certainly, one would hope to remove some or even most of these 'untrue' solutions by imposing constraints in the optimization framework. In a wider context, such ill-posed problems have been tackled in mathematical and statistical literature by invoking the sparsity condition (Tarantola, 2005; Kaipio and Somersalo, 2006). For example, compressed sensing in signal processing is facilitated by applying the sparsity constraint into a usually ill-posed optimization problem (Candes et al, 2006; Donoho 2006; Mascarenas et al, 2013). Interestingly, the sparsity condition is also applicable to damage identification problem. Specifically, in practical situation, damage occurrence is much more probable to affect only a small number of elements/segments in a finite element model of the structure. In other words, the unknown vector in the damage identification problem is sparse. However, few investigations have so far taken advantage of the inherent sparse nature of damage indices to address damage identification problems. Wang and Hao (2014) formulated a pattern recognition problem for damage identification and matched the pre-defined patterns following a compressed sensing-based scheme which uses sparsity properties and $l_1$ regularization on the unknown damage pattern vector. This approach cannot be easily extended to cases with many damage patterns. More recently, Huang et al (2017) adopted the sparse Bayesian learning for structural damage detection. Rather than providing a single estimate, sparse Bayesian learning provides a full posterior density function, which gives a sense of confidence of the approximation.

While it is intuitive to incorporate the sparsity condition into the aforementioned optimization problem, the formulation of an optimization problem (i.e., design objective with constrains) and the associated solution method are intertwined, especially when many unknown variables are involved. For the problem of damage identification, three questions arise, including 1) how to actually define sparsity condition mathematically, 2) how to integrate the sparsity condition into optimization, and 3) how to solve the optimization problem. In this paper, we develop a multi-objective optimization formulation that can be effectively solved by a deterministic global approach originated from the Dividing RECTangles (DIRECT) algorithm (Jones et al, 1993). The rationale is given as follows. The difference between model prediction in the parametric space and the measurement needs to be minimized for the sake of damage identification, which is one obvious objective function in the optimization problem. Although the sparsity condition can be included as a constraint, in this research we opt to treat it as a separate objective function to be minimized. First this can avoid the employment of weighting constant, which is generally ad hoc, in solving the usual constrained optimization problem. Second, this formulation naturally produces multiple solutions, which reflects the under-determined nature of the problem. Moreover, as will be shown comprehensively later in this paper, a proper choice of sparse regularization under the multi-objective formulation (i.e., treating the sparsity as the second objective function) has the potential of yielding a *small* set of solutions that fits better the true damage scenario. In comparison, if a single composite objective function is used, it yields one single solution which however may not capture the true damage at all. It is worth emphasizing that the DIRECT algorithm is particularly suitable in solving the multi-objective optimization problem formulated. Mathematically, it is a deterministic technique, and thus the results obtained are repeatable and conclusive without ambiguities. Only one algorithmic parameter, the number of function evaluations, is needed here. To enhance the computational efficiency, a new sampling/division scheme in the unknown parametric space is established. In this research, to demonstrate its effectiveness, this new framework of damage identification is applied to piezoelectric impedance-based SHM featuring high detection sensitivity but with large number of unknown variables.

The rest of the paper is organized as follows. In Section 2, we briefly outline the basic equations and background information of piezoelectric impedance-based technique, followed by the optimization problem formulation. The details of the sparse multi-objective DIRECT algorithm are presented in Section 3. In Section 4, the proposed method is evaluated through two benchmark damage scenarios. Further validations using experimental results are conducted in Section 5. Concluding remarks are given in Section 6.

## 2. PROBLEM FORMULATION

### 2.1 Piezoelectric impedance/admittance active sensing

In piezoelectric impedance/admittance-based damage detection, a piezoelectric transducer is attached to the host structure. Frequency-sweeping, harmonic excitation voltage is applied to the piezoelectric transducer, which induces structural oscillation. Owing to the two-way electro-mechanical coupling, the local structural oscillation in turn affects the electrical response of the transducer. As such, the impedance/admittance of the transducer is directly related to the impedance of the underlying structure. Indeed, the equations of motion of the coupled sensor-structure interaction system in the finite element form can be derived as (Wang and Tang, 2010),

$$\mathbf{M}\ddot{\mathbf{q}} + \mathbf{C}\dot{\mathbf{q}} + \mathbf{K}\mathbf{q} + \mathbf{K}_{12}Q = \mathbf{0} \tag{1a}$$

$$K_c Q + \mathbf{K}_{12}^{T}\mathbf{q} = V_{in} \tag{1b}$$

where $\mathbf{q}$ is the structural displacement vector, $\mathbf{M}$, $\mathbf{K}$ and $\mathbf{C}$ are the mass matrix, stiffness matrix and damping matrix, respectively, $\mathbf{K}_{12}$ is the vector indicating the electro-mechanical coupling due to piezoelectric effect, $K_c$ is the reciprocal of the capacitance of the piezoelectric transducer, $Q$ is the electrical charge on the surface of the piezoelectric transducer, and $V_{in}$ is the excitation voltage. Under harmonic excitation, Equations (1a) and (1b) can be transferred to frequency domain. The admittance of the piezoelectric transducer that is coupled to the host mechanical structure can be expressed as

$$Y(\omega) = \frac{\dot{Q}}{V_{in}} = \frac{\omega i}{K_c - \mathbf{K}_{12}^{T}(\mathbf{K} - \mathbf{M}\omega^2 + \mathbf{C}\omega i)^{-1}\mathbf{K}_{12}} \tag{2}$$

where $\omega$ is the excitation frequency and $i$ is the imaginary unit. While the admittance and the impedance are reciprocal with each other, here in this research without loss of generality we employ the piezoelectric admittance as the information carrier.

In discretized model-based structural health monitoring, structural damage is frequently assumed as local property change, e.g., local stiffness loss. We divide the host structure into $n$ segments and let $\mathbf{k}_{hj}$ be the stiffness matrix of the healthy, $j$-th segment. The stiffness matrix of the structure with damage is then expressed as

$$\mathbf{K}_d = \sum_{j=1}^{n}\mathbf{k}_{hj}(1-\alpha_j) \tag{3}$$

where $\alpha_j \in [0, 1]$ is the damage index indicating the stiffness loss in the $j$-th segment. For example, if the $j$-th segment suffers from damage that leads to a 20% stiffness loss, $\alpha_j = 0.2$. In practical situation, multiple segments may be subjected to damage or stiffness losses. Let $\boldsymbol{\alpha} = [\alpha_1, \cdots, \alpha_n]^T$ be the damage index vector.

Multiple vector elements in $\boldsymbol{\alpha}$ may be non-zero. Meanwhile, damage effect is reflected in the change of admittance of the piezoelectric transducer,

$$Y_d(\omega) = \frac{\dot{Q}_d}{V_{in}} = \frac{\omega i}{K_c - \mathbf{K}_{12}^T(\mathbf{K}_d - \mathbf{M}\omega^2 + \mathbf{C}\omega i)^{-1}\mathbf{K}_{12}} \qquad (4)$$

Typically, the measured admittance of the damaged structure is compared with the baseline admittance information in order to identify damage location and severity.

It can be observed from Equations (3) and (4) that the admittance change is not linearly dependent upon the damage index vector. When the size of damage is small, which is the usual situation, we can use Taylor series expansion to develop a linearized relationship between the admittance change and the damage index vector (Shuai et al, 2017). The admittance of the damaged structure can be expressed as,

$$Y_d(\boldsymbol{\alpha}) \approx Y(\boldsymbol{\alpha} = \mathbf{0}) + \sum_{j=1}^{n} \frac{\partial Y}{\partial \alpha_j}\bigg|_{\alpha_j=0} \alpha_j \qquad (5)$$

where

$$\frac{\partial Y}{\partial \alpha_j}\bigg|_{\alpha_j=0} = \omega i[k_c - \mathbf{K}_{12}^T(\mathbf{K} - \mathbf{M}\omega^2 + \mathbf{C}\omega i)^{-1}\mathbf{K}_{12}]^{-2}\mathbf{K}_{12}^T\left[\frac{\partial(\mathbf{K}_d - \mathbf{M}\omega^2 + \mathbf{C}\omega i)^{-1}}{\partial \alpha_j}\bigg|_{\alpha_j=0}\right]\mathbf{K}_{12} \qquad (6)$$

The change of admittance due to damage can then be written as,

$$\Delta Y(\omega) = Y_d - Y(\boldsymbol{\alpha} = \mathbf{0})$$
$$= \sum_{j=1}^{n}[\omega i(k_c - \mathbf{K}_{12}^T(\mathbf{K} - \mathbf{M}\omega^2 + i\omega\mathbf{C})^{-1}\mathbf{K}_{12})^{-2}\mathbf{K}_{12}^T(\mathbf{K} - \mathbf{M}\omega^2 + i\omega\mathbf{C})^{-1}\mathbf{k}_{hj}(\mathbf{K} - \mathbf{M}\omega^2 + i\omega\mathbf{C})^{-1}\mathbf{K}_{12}]\alpha_j \qquad (7)$$

Recall $\mathbf{k}_{hj}$ is the stiffness matrix of $j$-th ($j = 1, \ldots, n$) segment when the structure is intact. Equation (7) gives the relationship between the admittance change and the damage index at one specific excitation frequency point $\omega$. In impedance/admittance-based damage identification, while harmonic voltage excitation is used for active sensing, the frequency of the excitation is swept through a certain range that covers a number of structural resonances around which measurements are taken. Assume admittance values are measured at $m$ frequency points, $\omega_1, \cdots, \omega_m$. For each frequency point, Equation (7) holds whereas the damage index vector remains to be the same. We then have

$$\Delta \mathbf{Y} = \begin{bmatrix} \Delta Y(\omega_1) \\ \vdots \\ \Delta Y(\omega_m) \end{bmatrix} = \mathbf{S}_{m \times n}\boldsymbol{\alpha} \qquad (8)$$

where $\Delta \mathbf{Y}$, an $m$-dimensional vector, is the admittance changes measured at $m$ frequency points, $\boldsymbol{\alpha}$ is the $n$-dimensional vector of the unknown damage indices, and $\mathbf{S}_{m \times n}$ is the sensitivity matrix.

## 2.2 Inverse analysis formulated as an optimization problem with $l_0$ sparse regularization

The high detection sensitivity of impedance/admittance-based active sensing approach is built upon the high-frequency responses excited and measured. A very large number of finite elements are needed to establish the baseline model for credible prediction of high-frequency responses. As we divide the structure into segments to facilitate damage identification, the structural properties of each segment remains to be identified because each segment is susceptible of fault occurrence, which yields a large number of unknowns. On the other hand, structural faults generally manifest themselves around the peaks of the piezoelectric impedance/admittance curves only, which means the input measurement information is usually limited in practice. Moreover, it is mathematically difficult to select frequency points to ensure the full rank of the sensitivity matrix (that relates fault location/severity with measurement, shown in Equation (8)) even if the number of frequency points is large. Therefore, the inverse identification formulation typically is under-determined (Kim and Wang, 2014). Although one may apply artificial constraints to seek for such as least square solutions, these solutions may not reflect the true fault scenario.

Hereafter, we cast the inverse identification problem into an optimization framework. Let $\Delta \mathbf{Y}$ be the measured admittance change. The prediction of admittance change in the parametric space is denoted as $\Delta \hat{\mathbf{Y}} = \mathbf{S}\hat{\boldsymbol{\alpha}}$. Certainly, we need to minimize the difference between these two, i.e.,

$$\min \ \|\mathbf{S}\hat{\boldsymbol{\alpha}} - \Delta \mathbf{Y}\|_2 \tag{9}$$

where $\|\bullet\|_p$ denotes the $l_p$ norm defined as $\|\mathbf{x}\|_p = \left(\sum_i |x_i|^p\right)^{1/p}$. Equation (9) has a large number of local minima given that $\mathbf{S}$ is rank-deficient. It is worth noting that a true damage scenario in practical situation usually affect only a small number of segments. In other words, the damage index vector $\boldsymbol{\alpha}$ is sparse by nature. Here we introduce the sparse regularization by enforcing a sparse constraint on $\hat{\boldsymbol{\alpha}}$. Traditionally, an $l_2$ regularizer $\|\hat{\boldsymbol{\alpha}}\|_2$ has been used (Donoho, 2006), which, however, is not designed toward sparse solutions. As illustrated in Figure 1, larger $p$ tends to spread out the error more evenly among variables and return a non-sparse $\hat{\boldsymbol{\alpha}}$ with many nonzero elements. Thus, we are more interested in $l_0$ or $l_1$ norm. While $l_1$ norm is usually used in statistical and signal processing as a convex approximation of non-convex $l_0$ to improve computational efficiency (Davenport et al, 2011), it introduces a mismatch between the goal of itself and that of $l_0$ norm (Wipf and Rao, 2004). As such, we may fail to recover the maximally sparse solution regardless of the initialization. In damage identification, finding a solution of Equation (9) with the presence of modeling and measurement error cannot be simply solved as a convex optimization problem because it is indeed a multi-modal problem (having multiple local optima). Therefore, the convex approximation of $l_0$ would not be a necessity. Thus, the second objective function here is chosen as the minimization of the $l_0$ norm of $\hat{\boldsymbol{\alpha}}$, i.e.,

$$\min \|\hat{\boldsymbol{\alpha}}\|_0 \tag{10}$$

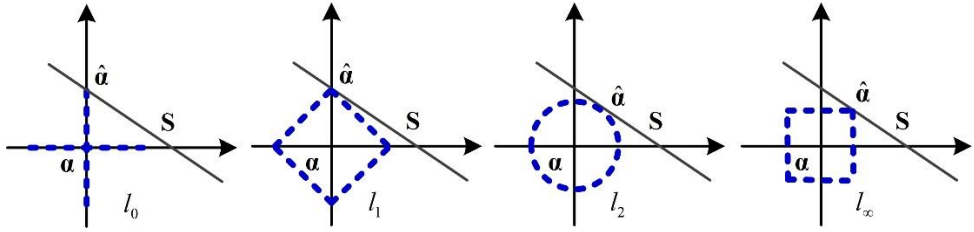

**Figure 1.** Approximation of point $\boldsymbol{\alpha}$ in $\mathbf{R}^2$ using $l_0$, $l_1$, $l_2$ and $l_\infty$ norm.

One way of handling the two objective functions selected above (Equations (9) and (10)) is to formulate a composite objective function, i.e. (Cao et al, 2016),

$$\min \ \|\mathbf{S}\hat{\boldsymbol{\alpha}} - \Delta\mathbf{Y}\|_2 + \lambda \|\hat{\boldsymbol{\alpha}}\|_0 \tag{11}$$

where $\lambda$ is the weighting factor. The choice of weights is usually ad-hoc since the relative importance of each objective is unknown. A more critical issue is that such a single objective optimization usually gives one single optimum which may or may not fit the true damage scenario at all. Indeed, mathematically, Equation (8) is oftentimes under-determined. While we know the damage index vector must be sparse, we generally do not know how sparse it is. In this research, we formulate a multi-objective optimization,

Find: $\hat{\boldsymbol{\alpha}} = \{\hat{\alpha}_1, \hat{\alpha}_2, ..., \hat{\alpha}_n\}$, $\quad \alpha^l \leq \hat{\alpha}_j \leq \alpha^u$, $\quad j = 1, 2, .n$

Minimize: $f_1 = \|\mathbf{S}\hat{\boldsymbol{\alpha}} - \Delta\mathbf{Y}\|_2$ and $f_2 = \|\hat{\boldsymbol{\alpha}}\|_0$ (12a, b)

where $\alpha^l$ and $\alpha^u$ are the lower bound and upper bound of the damage index. A fundamental advantage of this multi-objective optimization formulation is that it naturally yields a set of optimal solutions explicitly exhibiting the tradeoff between objectives, i.e., the Pareto front/surface (Deb et al, 2002). This fits exactly the under-determined nature of the damage identification problem, and provides identification results that can be used for further inspection and prognosis which is the actual procedure of performing SHM.

## 3. DAMAGE IDENTIFICATION USING MULTI-OBJECTIVE DIRECT
### 3.1 DIRECT algorithm

DIRECT algorithm is a deterministic global optimization technique originally formulated by Jones et al (1993). The algorithm facilitates optimization tasks by Dividing Rectangles, from which its name is derived. Inherently, as a deterministic approach, DIRECT converges faster than stochastic global techniques. Moreover, while the performance of most stochastic global optimization methods depend heavily on algorithmic parameters, DIRECT has much less tunable parameters. These features are desirable for tackling damage identification problems. The original DIRECT algorithm is summarized in four steps as follows.

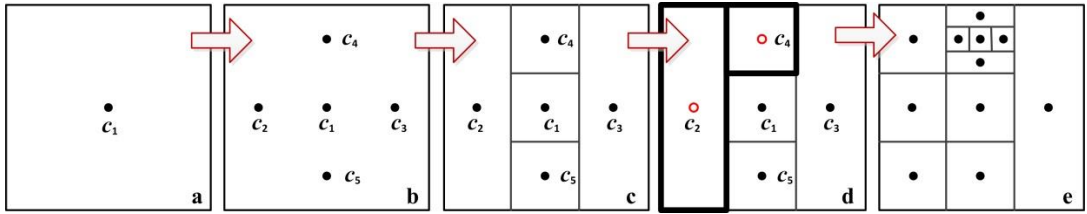

**Figure 2. Sampling and dividing of the decision space.**

*1) Normalize*. Normalize the decision domain to a unit cube/hyper-cube (Figure 2(a)).

$$\bar{\Omega} = \{x \in R^N : 0 \leq x_i \leq 1\}, \qquad i = 1, 2, ..., n \tag{13}$$

*2) Sample and divide*. Sample the center point $c_1$ of the cube/hyper-cube. Then evaluate the objective function values.

$$\mathbf{c}_1 \pm \mu \vec{e}_i, \qquad i = 1, 2, ..., n \tag{14}$$

where $\mu$ is one-third of the length of the cube, and $\vec{e}_i$ is the $i$-th unit vector in Euclidean space (Figure 2(b)). Dimension $i$ will be divided first if $i$ satisfies,

$$\arg\min_{i=1,2,...,n} \left( \min(f(\mathbf{c}_1 + \mu \vec{e}_i), f(\mathbf{c}_1 - \mu \vec{e}_i)) \right) \tag{15}$$

For example, if $\min(f(\mathbf{c}_2), f(\mathbf{c}_3)) < \min(f(\mathbf{c}_4), f(\mathbf{c}_5))$ (Figure 2(c)), the dimension along $\mathbf{c}_2, \mathbf{c}_3$ direction will be divided first.

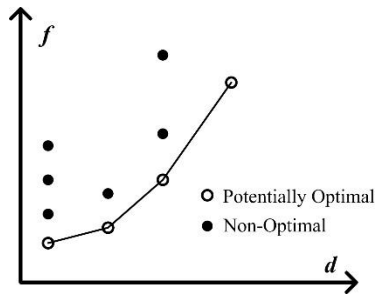

**Figure 3. Determine the potentially optimal rectangles.**

*3) Determine the potentially optimal rectangles.* A rectangle *j* is said to be potentially optimal if some rate-of-change constant $K > 0$ and a positive constant $\varepsilon$ exists satisfies,

$$f(\mathbf{c}_j) - Kd_j \leq f(\mathbf{c}_i) - Kd_i, \quad \text{for } \forall i \tag{16a}$$

$$f(\mathbf{c}_j) - Kd_j \leq f_{\min} - \varepsilon |f_{\min}| \tag{16b}$$

where $d_j$ measures the size of the *j*-th rectangle. In the original formulation of DIRECT, *d* is the distance from rectangle center *c* to the vertex. In Figure 2(d), the rectangles containing $\mathbf{c}_2$ and $\mathbf{c}_4$ are potentially optimal and thus are further divided as shown in Figure 2(e).

*4) Iterate.* Repeat Step 2 and Step 3 until maximum number of evaluation is reached. In Figure 2, a 2D example is provided. Higher dimensional cases could be divided and sampled following the similar procedure (Jones et al, 1993; Gablonsky, 2001).

### 3.2 Multi-objective DIRECT

The original DIRECT is designed for single-objective optimization. In this sub-section, aiming at solving the multi-objective optimization problem defined in Equation (12), we devise a multi-objective DIRECT approach. There have been some recent multi-objective variants of DIRECT (Wang et al, 2008; Al-Dujaili and Suresh, 2016; Wong et al, 2016). In the approach suggested by Wang et al (2008), rank strategy was used to rank the rectangles into fronts in terms of their objective values, and then $f(c)$ in Equation (16) was replaced with the rank to identify the potentially optimal rectangles. However, this approach only explores the rectangles on lower-right convex front (Figure 4) and overlooks a portion of rectangles that may lead to optimal solutions. Al-Dujaili and Suresh (2016) incorporated *d* as an additional objective value and obtained the potentially optimal rectangles by calculating the non-dominated Pareto front. One drawback of such a technique is that it expands the definition of potentially optimal to the extent where most rectangles are considered potentially optimal while only a small portion of them are rectangles of interest. Wong et al (2016) replaced *R* in Figure 4 with the hypervolume indicator and selected rectangles on the upper right Pareto front in the *hypervolume-d* plane. This approach is computationally prohibitive because hypervolume needs to be calculated iteratively, whereas hypervolume calculation is a computationally expensive NP-hard task itself.

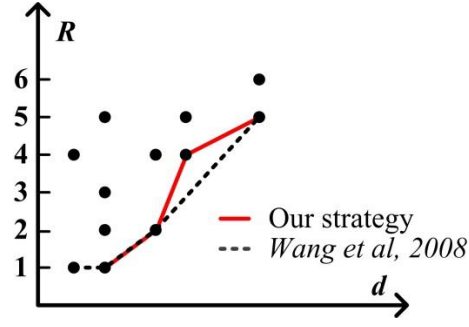

**Figure 4. Determine the potentially optimal rectangles based on rank index $R$ ($d$: rectangle size).**

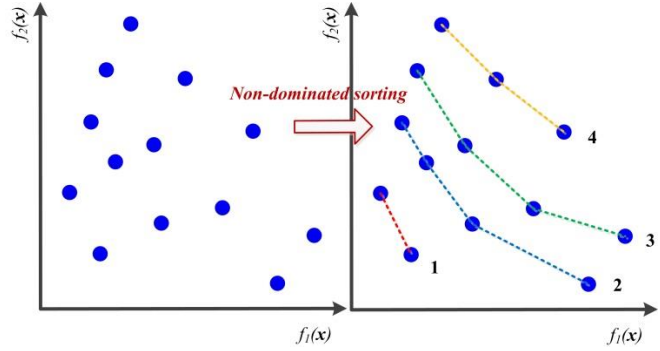

**Figure 5. Rank based on objective values.**

In this research, we employ a new strategy to fulfill the multi-objective capability and meanwhile address the aforementioned shortcomings. To determine the potentially optimal rectangles, the rectangles are ranked into fronts/surfaces using non-dominated sorting (Deb et al, 2002) as illustrated in Figure 5. By definition, the first front has a rank index of 1, the second front has a rank index of 2, and so on. Then the potentially optimal rectangles can be attained by projecting the rectangles to $R$-$d$ plane and extract the rectangles on the lower right Pareto front as shown in Figure 4. In the case of $h$-objective minimization, a rectangle $j$ is said to be potentially optimal if

$$R\big(R\big(f_1(c_j),\ f_2(c_j),...,f_h(c_j)\big),-d\big)=1 \tag{17}$$

where $R(\cdot)$ is an non-dominated sorting operator that returns the rank index. In the original implementation of DIRECT, $d$ is the distance measured between the rectangle center and vertex. In this research, $d$ is set as the length of the longest side of the rectangle such that the algorithm groups more rectangles at the same size (Finkel, 2005). Compared to the original DIRECT algorithm, the second condition (Equation 16(b)) is omitted because by considering the rectangles on lower right Pareto front, the rectangles on lower right convex hull that have the same $R$ value but smaller $d$ value will be naturally eliminated, which entertains the same effect of Equation 16(b). In a prior study, the proposed strategy has been compared to the above-mentioned three contemporary multi-objective DIRECT techniques, and exhibits favorable performance (Cao et al, 2017).

### 3.3 DIRECT for sparse exploitation

In this sub-section, we discuss the advantages of using DIRECT algorithm and the sigmoid transformation for finding sparse solutions for damage identification problem. For an *n*-dimensional optimization problem, the center $\mathbf{c_1}$ of the normalized decision space $\bar{\Omega}$ can be represented by a vector $[c_{1,1}, c_{1,2}, c_{1,3} \ldots c_{1,n}]^\mathrm{T}$. The neighboring decision space of $\mathbf{c_1}$ will be explored first since DIRECT divides along each dimension concurrently to identify the possibility of improvement (Figure 6). Therefore, if the optimal solution stays within the vicinity of the initial center, the solution can be approximated in certain number of steps due to the deterministic nature of DIRECT. In other words, if the sparsity of the solution is known *a priori*, the computational efficiency can be improved tremendously by adjusting the decision space so the center of the normalized decision space $\bar{\Omega}$ is a 0-sparse vector, i.e., zero vector. This is especially crucial when dealing with high dimensional damage identification problems where the computational cost for a global optimization is prohibitive. Hence, incorporating the fact that the damage index vector $\boldsymbol{\alpha}$ must be sparse into the problem formulation not only is necessary but also can facilitate the computationally efficient implementation of multi-objective DIRECT.

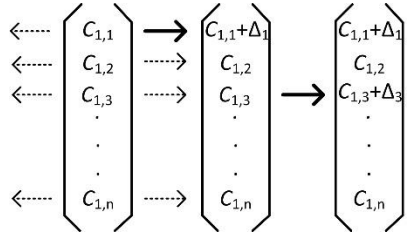

**Figure 6. Neighboring exploration of DIRECT.**

Recall Equation (13). The decision space of an *n*-dimensional damage identification problem can be denoted as $\alpha^l \leq \hat{\alpha}_j \leq \alpha^u$, $j = 1, 2, ..., n$. In order to make $[c_{1,1}, c_{1,2}, c_{1,3} \ldots c_{1,n}]^\mathrm{T}$ a zero vector, we need to have $-\alpha^l = \alpha^u$. As explained in Section 2, $\alpha_j \in [-1, 1]$ is the damage index indicating the stiffness loss (+) or gain (-) in one segment. Naturally, $\alpha^l$ and $\alpha^u$ can be chosen as -1 and 1, respectively. However, actual damage is usually very small stiffness losses, and thus the choice of -1 and 1 may lead to unnecessary computational burden. To direct the algorithm to search within a small neighboring space of a sparse vector, we adopt the following sigmoid function,

$$\text{sig}(x) = \frac{1}{1+e^{-bx}} \tag{18}$$

where $x \in (-\infty, \infty)$ and $\text{sig}(x) \in (0,1)$. As shown in Figure 7(a), the parameter $b$ controls the shape of the curve; larger $b$ indicates faster convergence to 0 and 1. With the introduction of the sigmoid function, an $n$-dimensional decision space with unknown bounds can be transferred to a cube/hypercube bounded by $\mathbf{0} \in \mathbf{R}^n$ and $\mathbf{1} \in \mathbf{R}^n$. The DIRECT search will therefore be conducted within $(\mathbf{0}, \mathbf{1})$ range with the center point at $\mathbf{0.5} \in \mathbf{R}^n$. As illustrated in Figure 7(b), the search will have a better focus on the vicinity of zero vector by using larger $b$. The inverse sigmoid function is defined as,

$$x = \frac{1}{b}(\ln(\text{sig}(x)) - \ln(1 - \text{sig}(x))) \tag{19}$$

In this research, $b$ is set to be 1,000 so that it agrees with our understanding of the model and helps to facilitate sparse exploitation without the need to define hard bounds.

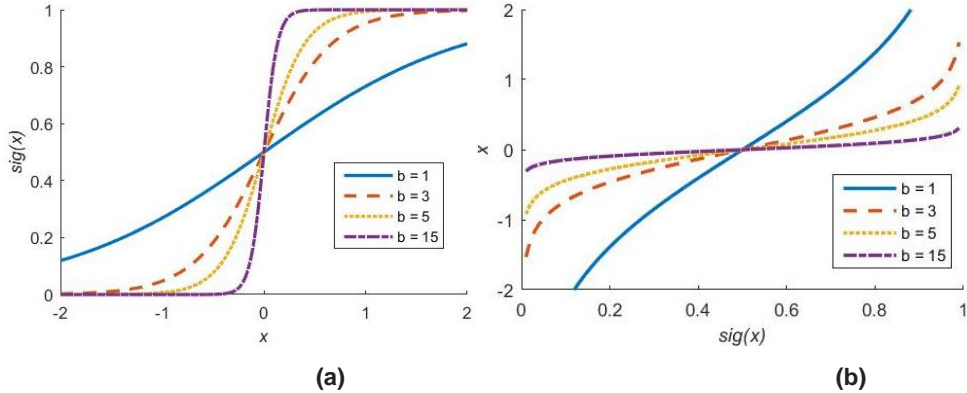

(a)            (b)

**Figure 7. Sigmoid and inverse sigmoid curves with b=1, 3, 5 and 15.**

Finally, for an $h$-objective optimization problem, the pseudo-code of the sparse multi-objective DIRECT is presented as follows.

Algorithm **sparse multi-objective DIRECT**

---

1: Transfer the search space to a unit cube/hypercube through sigmoid function

2: Evaluate $h$-objectives $\mathbf{F}(\mathbf{c}_1) = \{f_1(\mathbf{c}_1), f_2(\mathbf{c}_1), ..., f_h(\mathbf{c}_1)\}$; set $\{\mathbf{\alpha}_{\min}\} = \varnothing$, $k = 1$, and $k_{\max}$ = *max no. of evaluations*

3: **while** $k < k_{max}$

4:    Identify the set **S** of potentially optimal rectangles using Equation (17)

5:    **while** $\mathbf{S} \neq \varnothing$

6:       Take $j \in \mathbf{S}$

7:       Sample new points, evaluate **F** at the new points and divide the rectangle $j$ following step *2)*

8:       $k := k + \Delta k$

9:       $\mathbf{S} := \mathbf{S} \setminus \{j\}$

10:   **end while**

11: **end while**

12: **for** i = 1 to $k$

13:    **if** $R(\mathbf{F}(\mathbf{c}_i)) = 1$        $\{\boldsymbol{\alpha}_{\min}\} := \{\boldsymbol{\alpha}_{\min}\} \cup sig^{-1}(\mathbf{c}_i)$

14: **end for**

15: **return** $\{\boldsymbol{\alpha}_{\min}\}$

---

The returned vector set $\{\boldsymbol{\alpha}_{\min}\}$ is the solution set obtained from Equation (12). A posterior articulation can be performed by omitting $l_0$ norm $f_2 = \|\hat{\boldsymbol{\alpha}}\|_0$ and then finding damage identification result $\hat{\boldsymbol{\alpha}}$ in $\{\boldsymbol{\alpha}_{\min}\}$ with smaller residue $f_1 = \|\mathbf{S}\hat{\boldsymbol{\alpha}} - \Delta\mathbf{Y}\|_2$. In line 13, $R(\mathbf{F}(\mathbf{c}_i)) = 1$ indicates that the solution has a rank index of 1, i.e., the solution belongs to the Pareto surface. We have now outlined the approach to be used. Figure 8 depicts the overall structure of the approach adopted for damage identification where the dashed boxes are the steps that promote sparsity of the solution.

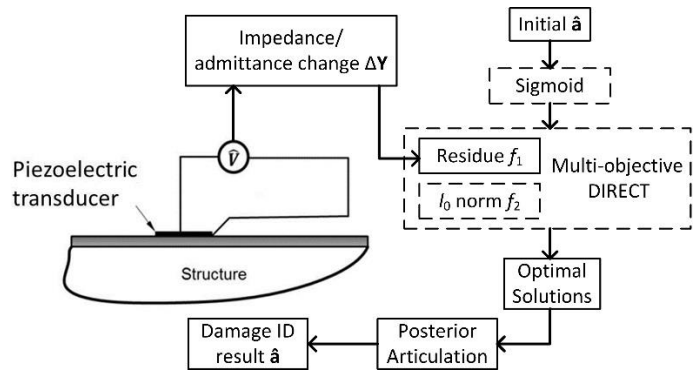

**Figure 8. Damage identification using piezoelectric impedance/admittance measurement: overview.**

# 4. NUMERICAL CASE STUDIES

In this section, we carry out comprehensive numerical case studies. We consider an aluminum cantilever plate with the following properties: length 0.561 m, width 0.01905 m, thickness 0.004763 m, density 2700 kg/m$^3$, and Young's modulus 68.9 GPa. The plate is subjected to the fixed-free boundary condition along the length-wise direction. A piezoelectric transducer is attached to this plate at location that is 0.18 m from the fixed end. The transducer has the following properties: length 0.015 m, width 0.01905 m, thickness 0.015 m, Young's moduli $Y_{11} = 86$ GPa and $Y_{33} = 73$ GPa, density 9500 kg/m$^3$, piezoelectric constant $-1.0288 \times 10^9$ V/m, and dielectric constant $\beta_{33} = 1.3832 \times 10^8$ m/F. The finite element model of the plate has 11,250 ($375 \times 15 \times 2$) 20-node hexahedron elements. It is further divided into 25 segments for damage identification indicating 25 possible damage locations. The segments are evenly divided along the length direction of the plate. In damage detection using admittance measurements, the admittance changes due to damage occurrence are most evident around the resonant peaks. Without loss of generality, we acquire the admittance change information around the plate's 14$^{th}$ (1893.58Hz) and 21$^{st}$ (3704.05Hz) natural frequencies. By keeping the linearly dependent rows of the sensitive matrix, a total of 21 useful measurements around the two natural frequencies are employed in the inverse analysis. As discussed in Section 1, the problem is under-determined. The linearization approximation (Equation (5)) also introduces error. For both test cases presented below, the maximum number of function evaluation $k_{max}$ is set as 100,000.

Table 1. Optimal results: numerical test case 1.

| Optimal Results | Residue ($f_1$ in Equation (12)) | # of Damage ($f_2$ in Equation (12)) |
|---|---|---|
| 1 | 1.8223e-07 | 1 |
| 2 | 1.1771e-10 | 2 |
| 3 | 1.1770e-10 | 3 |
| 4 | 1.1763e-10 | 4 |
| 5 | 1.1704e-10 | 5 |
| 6 | 1.1698e-10 | 6 |
| 7 | 1.1698e-10 | 7 |

We perform the first test with two randomly selected damage locations on the 9$^{th}$ and 21$^{st}$ segment with severities $\alpha_9 = 0.03$ (3% stiffness loss) and $\alpha_{21} = 0.05$ (5% stiffness loss) respectively. As shown in Table 1, multiple optimal solutions are produced by the sparse multi-objective DIRECT optimization. The solutions obtained exemplify the tradeoff between residue (Equation (12a)) and the number of damage locations (Equation (12b)). In other words, the solution with smaller residue may have more predicted damaged segments. As shown in Figures 9(b)-(g), optimal results 2 to 7 are relatively similar to each other in terms of identified results. In genetic algorithm description, they are considered to belong to the same niche. Consider optimal results 2 (Figure 9(b)) and 7 (Figure 9(g)) as example. The latter has two major damages (similar to optimal result 2) and six more negligible damages that reflect the error introduced by linear approximation and analytical modeling. Since the number of damage is unknown in practice, we could consider the niches of optimal result 1 and 2 as possible candidates. However, as illustrated in Figure 10, optimal result 2 matches the input data better than optimal result 1 in terms of residue value. Thus, the niche of optimal result 2 is more preferable, which indeed agree with the true damage scenario.

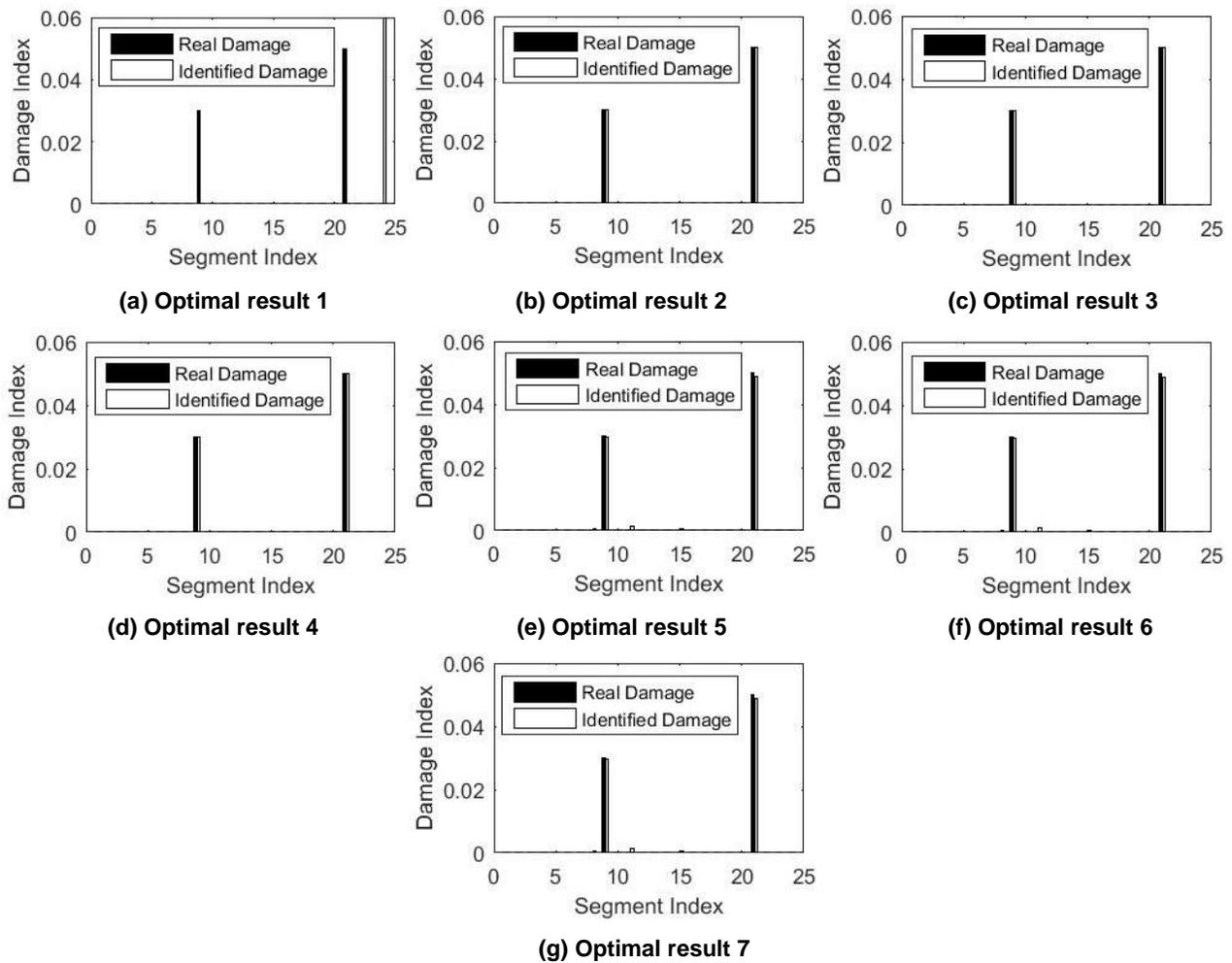

**(a) Optimal result 1**  **(b) Optimal result 2**  **(c) Optimal result 3**
**(d) Optimal result 4**  **(e) Optimal result 5**  **(f) Optimal result 6**
**(g) Optimal result 7**

**Figure 9. Numerical test case 1: real damage scenario compared with predicted damage scenarios.**

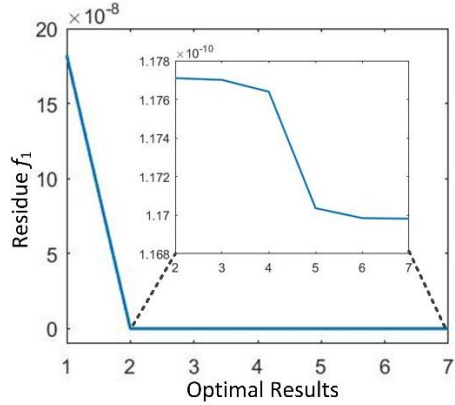

**Figure 10. Numerical test case 1: Residue values of optimal results.**

**Table 2. Optimal results: numerical test case 2.**

| Optimal Results | Residue ($f_1$ in Equation (12)) | # of Damage ($f_2$ in Equation (12)) |
|---|---|---|
| 1 | 8.2871e-07 | 1 |
| 2 | 2.4805e-08 | 2 |
| 3 | 7.0532e-13 | 3 |
| 4 | 2.0381e-13 | 4 |
| 5 | 1.9074e-13 | 5 |
| 6 | 2.9357e-14 | 6 |

Next we carry out the second case study with damage randomly selected at the $6^{th}$, $13^{th}$, and $20^{th}$ segments with severities of 0.04, 0.02 and 0.03 respectively. In this case, the number of damage is increased. Similarly, a set of optimal results consisting of the best solution for each possible number of damage are obtained and listed in Table 2. It is observed in Figure 11 that four out of six predictions successfully locate the damage and approximate the severities with excellent accuracy. Even though optimal results 2 and 3 are also plausible, by scrutinizing the residue values (Figure 12), we can find that solutions within the niche of optimal result 3 have smaller residue values, which indicate that they match with the input information better. For the proposed approach, when $k_{max}$ is large enough, each optimal result obtained is the best result amongst its type in terms of the number of damage locations. Subsequently, even though such number is generally unknown during the inverse identification, the true damage scenario would be identified and the corresponding result will be included in the optimal set as long as the error is well-controlled and the problem is identifiable. In general, a posterior articulation can then be performed by selecting the results with smaller residue values. In certain situations when we know *a priori* that considerable modeling error and measurement noise are present, we can further look into results with relatively large residues. In any case, the proposed approach yields a small set of probable damage identification results with clear quantification of performance trade-offs (in terms of the number of damage locations and residue values). A final decision of damage identification can be reached by combining the piezoelectric admittance-based result with further inspection and/or additional monitoring schemes.

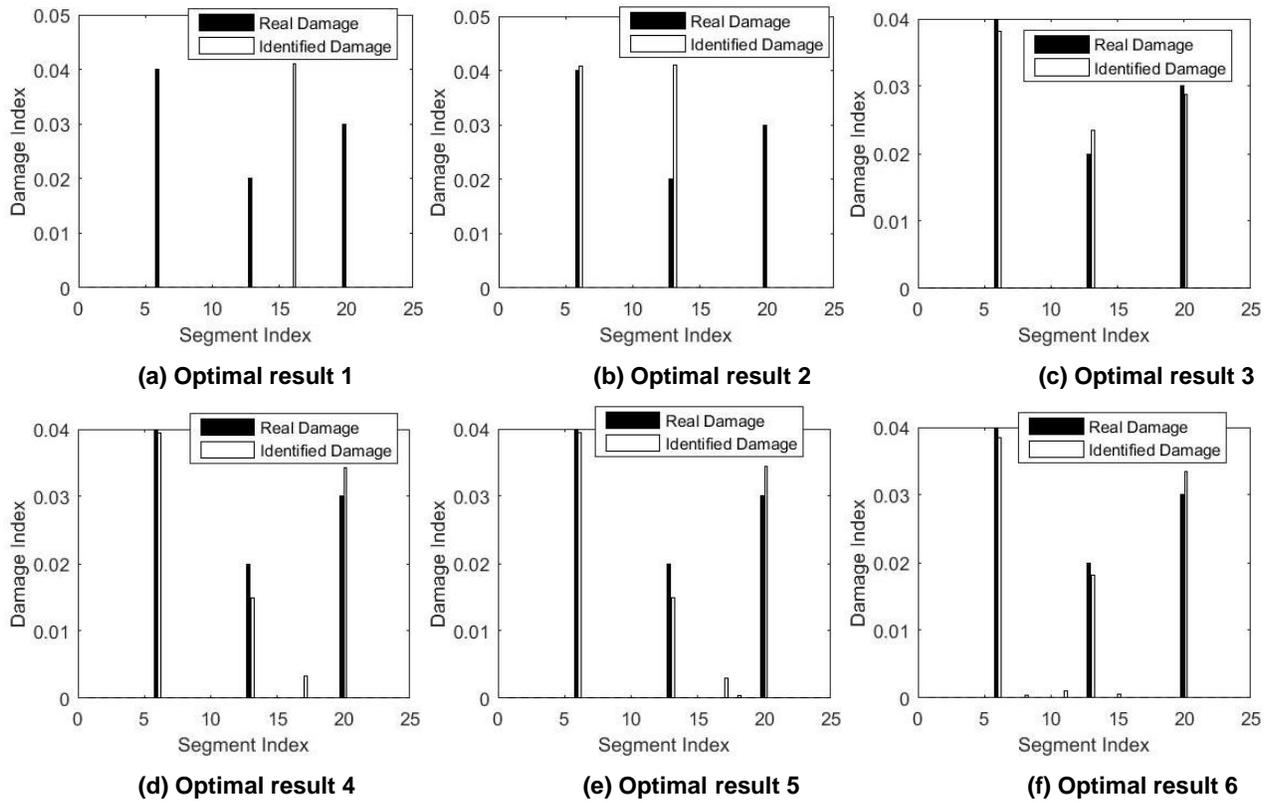

**Figure 11. Numerical test case 2: real damage scenario compared with predicted damage scenarios.**

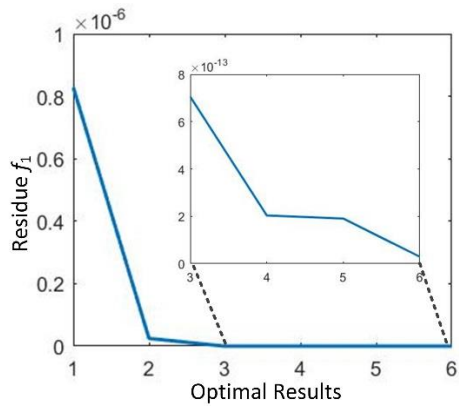

**Figure 12. Numerical test case 2: Residue values of optimal results.**

We further use the numerical study to elucidate the influence of the choice of sparsity regularizer. By changing $f_2 = \|\hat{\boldsymbol{\alpha}}\|_0$ in Equation (12b) to $f_2 = \|\hat{\boldsymbol{\alpha}}\|_1$ and applying sparse multi-objective sparse DIRECT, we conduct the comparison between the performance of $l_0$-norm and $l_1$-norm for the second numerical case study. When $l_1$-norm is employed, 1,344 optimal solutions are generated (Figure 13(b)) compared to 6 by $l_0$-norm (Figure 13(a)). Using $l_1$-norm poses challenge of performing a posterior articulation to start with. While for $l_0$-norm, even though it is not assured which candidate reflects the true damage state, the results obtained provide guidance for further inspection and decision making, because the number of solutions is typically much smaller than the number of segments. Moreover, the quality of the solutions obtained by optimization using $l_1$-norm is worse than that of $l_0$-norm. As shown in Figure 14, the mean value of the optimal set of $l_0$-norm is more accurate and closer to the true damage state than $l_1$-norm as a whole. Additionally, the best solution in terms of the residue value, which corresponds to the point on the upper left corner in Figure 13, is 2.7343e-12 for $l_1$-norm, which is much larger than that of $l_0$-norm (2.9357e-14). As discussed in Section 2.1, $l_1$-norm serves as a convex approximation of $l_0$-norm so that the maximally sparse solutions are recovered with additional error.

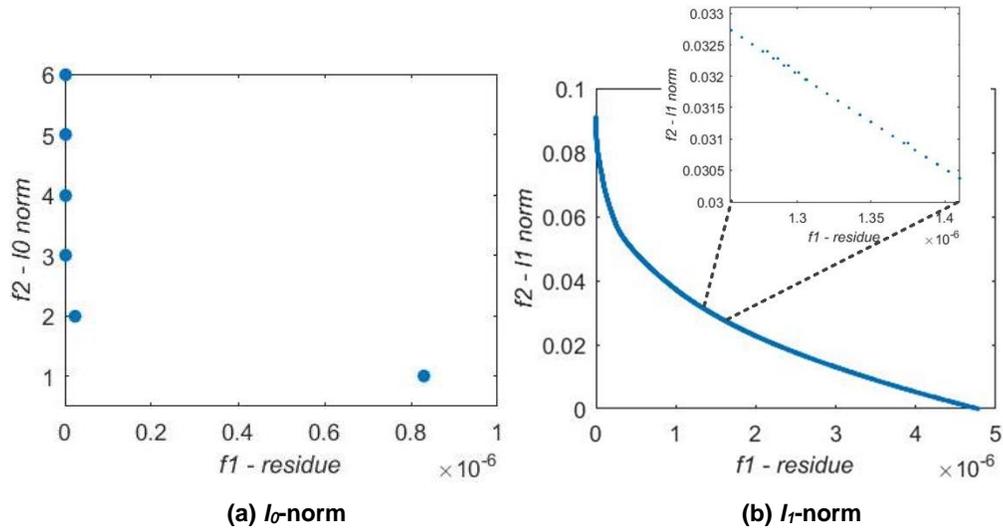

(a) $l_0$-norm  (b) $l_1$-norm

**Figure 13. Optimal results of numerical test 2.**

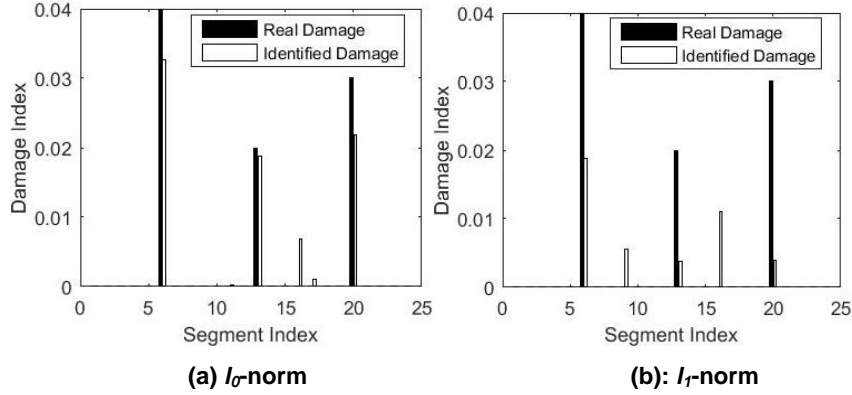

(a) $l_0$-norm          (b): $l_1$-norm

**Figure 14. Mean values of the predictions in the optimal sets.**

## 5. EXPERIMENTAL VALIDATIONS

In this section, experimental cases using actual piezoelectric admittance measurements are carried out. The experimental setup, geometry measures and material parameters are consistent with those in the numerical cases analyzed in Section 4. To obtain the piezoelectric admittance, a resistor of $100\,\Omega$ is connected in series with the transducer to measure the voltage drop, which, together with the current in the circuit, yields the admittance information. A signal analyzer (Agilent 35670A) with a source channel is employed. The source channel is used to generate sinusoidal voltage sent to piezoelectric transducer ($V_{in}$), and the output voltage across the resistor is recorded ($V_{out}$). Hence, the admittance can be obtained,

$$Y_{exp} = \frac{I}{V_{in}} = \frac{V_{out}}{R_s V_{in}} \tag{20}$$

Meanwhile, we need to calibrate the finite element model of the healthy structure to match the experimental setup to minimize the modeling error usually induced by non-perfect boundary conditions. We update the boundary conditions through numerical optimization by identifying the stiffness values at the fixed edges so the errors of the natural frequencies between the measurements and model predictions are minimized (Shuai et al, 2017). Comparisons between measured admittances and those generated by the numerical model before and after model updating are illustrated in Figure 15. The model updating yields very good match with the healthy baseline.

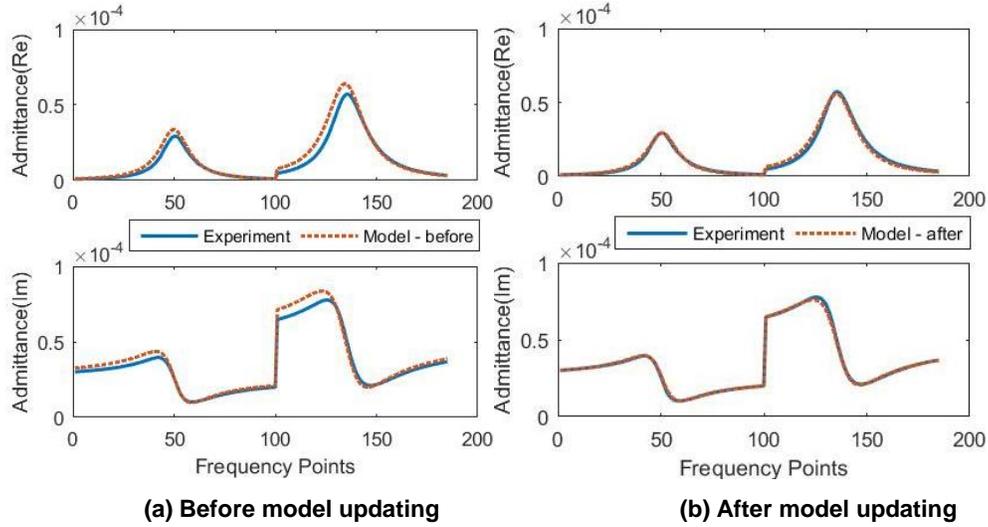

**(a) Before model updating**        **(b) After model updating**

**Figure 15. Admittance measured by experiment vs. admittance generated by numerical model under healthy condition.**

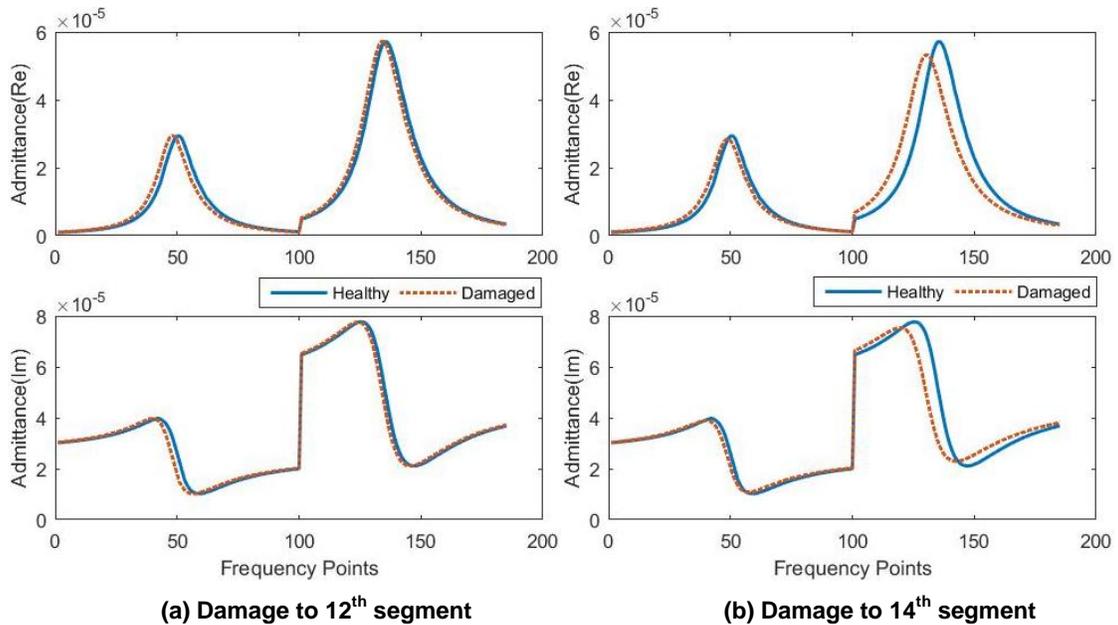

**(a) Damage to 12$^{th}$ segment**        **(b) Damage to 14$^{th}$ segment**

**Figure 16. Change of admittances (real and imaginary parts) caused by damage.**

In order to minimize unwanted variations and uncertainties in the experiment, instead of disassembling and cutting the plate to reduce the local stiffness, we employ small added masses to emulate the damage occurrence. This can facilitate easy change of damage scenarios without altering the experimental fixture. Practically, an added mass results in the change of admittance signatures equivalent to a local stiffness reduction. In the first experimental case, a 0.6 g mass is attached onto the plate at location corresponding to the 12$^{th}$ segment in the model, which causes admittance change equivalent to a 0.16% local stiffness loss (Figure 16(a)). In the second experiment, the same mass is attached onto the 14$^{th}$ segment of the plate which is equivalent to a 0.28% local stiffness loss (Figure 16(b)). Similar to numerical test cases, the admittances are measured around the 14$^{th}$ resonant frequency at 100 frequency points and the 21$^{st}$ resonant frequency at 85 frequency points. For both cases, the maximum number of function evaluation $k_{max}$ is set as 100,000.

We first implement the new approach to tackle the case where the 12$^{th}$ segment is subjected to 0.16% stiffness loss. Three optimal results are acquired in Table 3 utilizing sparse multi-objective DIRECT. Unlike numerical case studies, there is no magnitude difference between the optimal results in terms of residue. While there is only linear approximation error in numerical case studies, model error and measurement error are inevitable in experiments, which in turn blur the boundary of solutions that reflect the admittance shift caused by damage or by errors. Thus, we consider each to be a candidate. As illustrated in Figure 17(a), optimal result 1 in our optimal set has the best accuracy. Nevertheless, given no prior knowledge of the number of damaged segments, optimal results 2 and 3 are also likely and they explain the as-measured date even better. Ideally, if the experimental setup and the numerical model have complete agreement, the new approach will converge to a single solution that matches the true damage scenario perfectly. But in practice, the convergent solution could be different because of error, and it is computationally intensive to merely minimize the residue without sparse relaxation. However, by using the proposed approach, not only the computational burden can be alleviated, but also, a small set of possible solutions containing the ones that are in close proximity of the true damage scenario are collected. This provides the foundation for subsequent inspections.

To better elaborate the significance of the proposed approach, Figure 18 provides an example of how decision space is sampled by the multi-objective DIRECT algorithm in both the sigmoid space and the Euclidean space. Only 3 dimensions out of 25, which correspond to the 11$^{th}$, 12$^{th}$ and 13$^{th}$ segment, are selected for visualization. It is shown that the "damaged" dimension 12 is intensively sampled while the samples in dimensions of $\alpha_{11}$ and $\alpha_{13}$ mostly stay around zero in Euclidean space.

**Table 3. Optimal results: experimental case 1**

| Optimal Results | Residue ($f_1$ in Equation (12)) | # of Damage ($f_2$ in Equation (12)) | Damage Scenarios |
|---|---|---|---|
| 1 | 8.3545e-11 | 1 | 0.17% at 12$^{nd}$ |
| 2 | 8.1254e-11 | 2 | 0.18% at 10$^{th}$; 4.75e-2% at 23$^{rd}$ |
| 3 | 8.1156e-11 | 3 | 0.12% at 10$^{th}$; 0.13% at 17$^{th}$; 4.94e-3% at 18$^{th}$ |

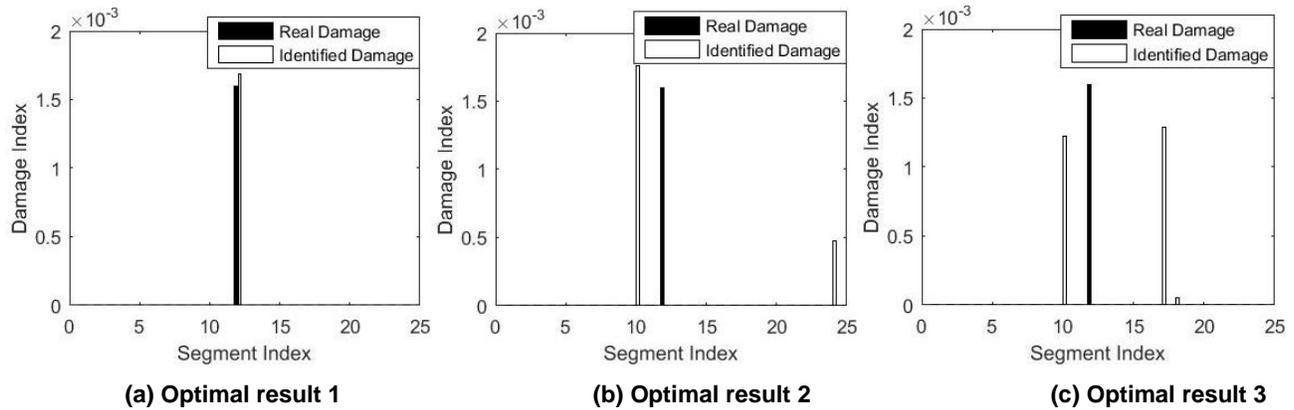

**(a) Optimal result 1**  **(b) Optimal result 2**  **(c) Optimal result 3**

**Figure 17. Experimental case 1: real damage scenario compared with predicted damage scenarios.**

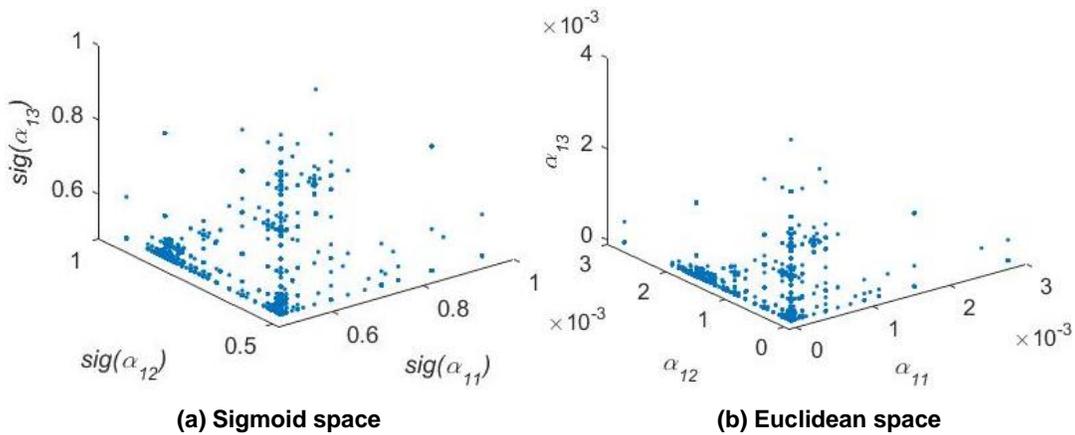

**(a) Sigmoid space**  **(b) Euclidean space**

**Figure 18. Experimental case 1: Multi-objective DIRECT Sampling.**

Next, we perform the second experimental case study where the $14^{th}$ segment is subjected to 0.28% stiffness loss. Table 4 lists four possible damage scenarios yeilded by the proposed approach where the optimal result 1 agrees with the true damage scenario. As shown in Figure 19, optimal results 2 to 4 are very similar to each other and thus can be regarded as the niche of optimal result 2 altogether. Without knowing the number of damaged segments, the niche of optimal result 1 and optimal result 2 are both possible candidates. As shown in Figure 20, solutions with more damages interpret the data better. However, better interpretation does not necessarily mean better solution. In fact, the tradeoff between the residue and the $l_0$-norm can be regarded as the tradeoff between over-fitting and under-fitting. While errors are expected in SHM systems, the optimal set provided by the proposed approach actually unfolds the best options for further inspections. Take Figure 19(a) for example. If we want the result to reflect merely the main essence of the data, optimal result 1 on the lower right corner should be selected. On the contrary, optimal result 3 on the upper left corner should be chosen if we want the result to fit the data perfectly. This kind of explicit trade-offs yielded by the proposed approach can give clear guidance for further inspection, which matches with the typical procedure of inspection and maintenance in practice.

**Table 4. Optimal results: experimental case 2.**

| Optimal Results | Residue ($f_1$ in Equation (12)) | # of Damage ($f_2$ in Equation (12)) | Damage Scenarios |
|---|---|---|---|
| 1 | 2.4601e-09 | 1 | 0.3% at $14^{th}$ |
| 2 | 2.3634e-09 | 2 | 0.23% at $14^{th}$; 6.68e-2% at $18^{th}$ |
| 3 | 2.3634e-09 | 3 | 2.74e-4% at $10^{th}$; 0.23% at $14^{th}$; 6.65e-2% at $18^{th}$ |
| 4 | 2.3634e-09 | 4 | 2.74e-4% at $10^{th}$; 0.23% at $14^{th}$; 6.65e-2% at $18^{th}$; 9.14e-5% at $24^{th}$ |

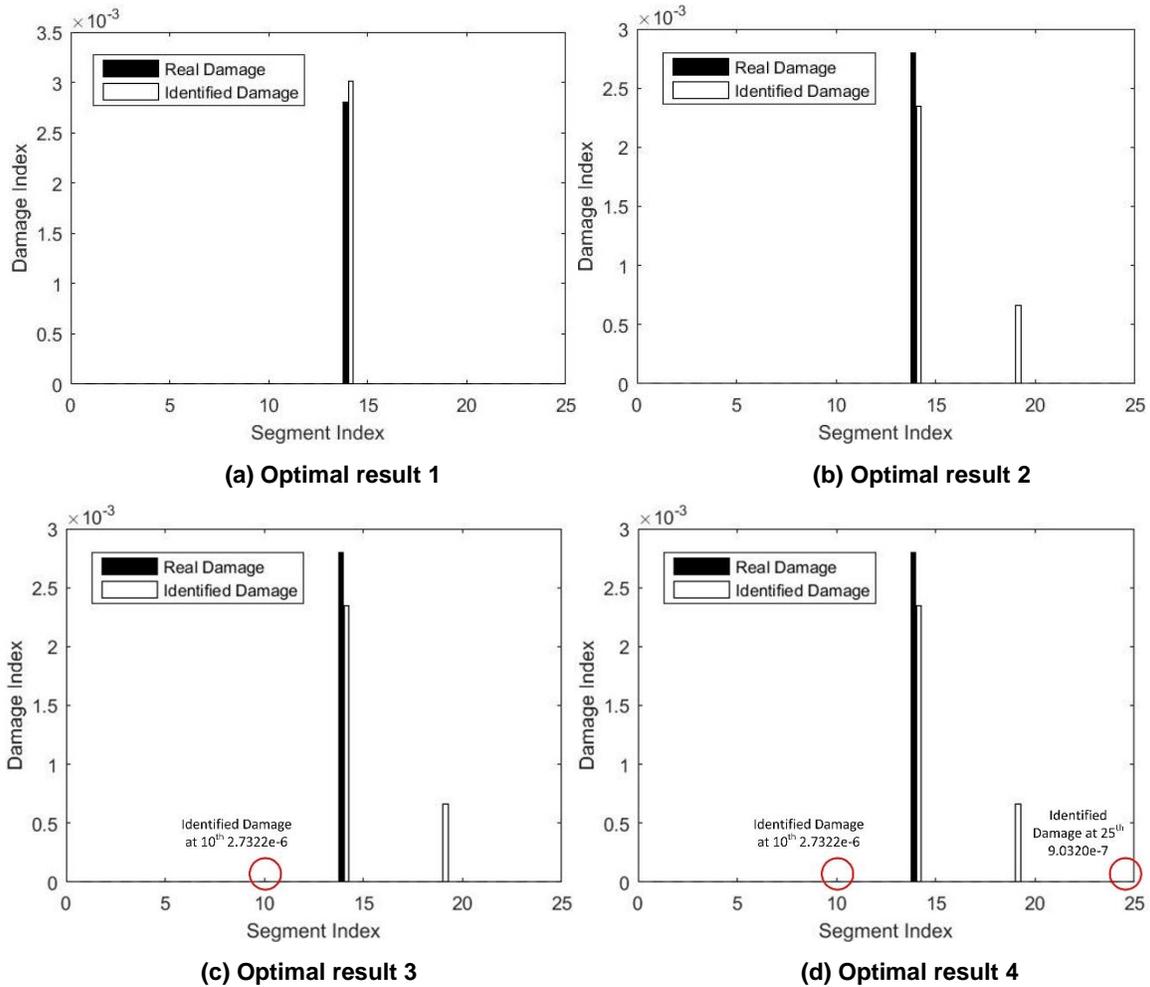

Figure 19. Experimental case 2: real damage scenario compared with predicted damage scenarios.

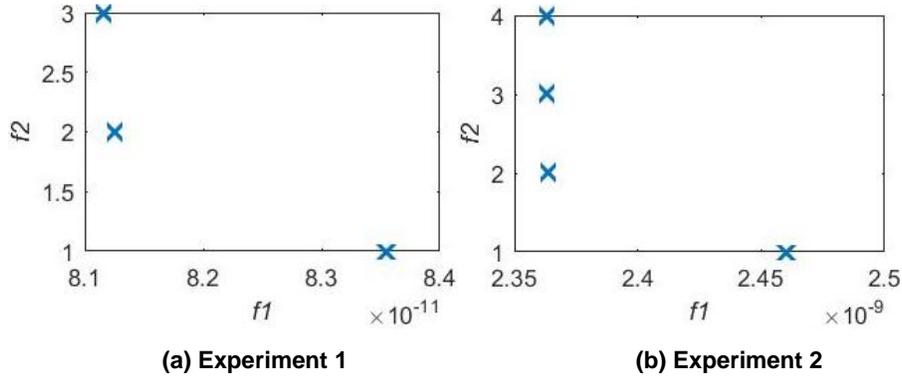

(a) Experiment 1  (b) Experiment 2

**Figure 20. Optimal results of experimental cases.**

## 6. CONCLUDING REMARKS

This research presents an effective approach for structural damage identification utilizing piezoelectric admittance measurements.  To address the fundamental challenge posed by the under-determined problem formulation that is rooted in the nature of high-frequency actuation/sensing, we cast the damage identification problem into a 2-objective optimization problem.  The optimization problem is then tackled by a newly devised sparse multi-objective DIRECT algorithm.  While the number of unknowns is very large in damage identification using high-frequency response measurements, in practical situations damage only affects a small number of segments.  The proposed approach exploits the sparsity of the solution through $l_0$-norm minimization and sigmoid transformation.  This not only reduces the number of solutions but also alleviates the computational burden for a global optimization.  The numerical tests and experiment validations demonstrate that the proposed algorithm is capable of obtaining a small set of high quality solutions that cover the true damage scenario.  Instead of seeking for a deterministic solution which could be very different from the actual damage, this proposed approach utilizes SHM measurements to identify probable damage locations and severities that can be further examined/inspected.  Such a procedure fits the practical SHM scheme.

## ACKNOWLEDGMENTS

This research is supported by the Air Force Office of Scientific Research under grant FA9550-14-1-0384.